\documentstyle[aps,epsf]{revtex} 

\def\be{\begin{equation}}
\def\ee{\end{equation}}   
\def\bea{\begin{eqnarray}}
\def\eea{\end{eqnarray}}

\begin{document}

\begin{flushright}
To my grandmother Perunika\\
D.K.
\end{flushright}

\vfill
\begin{center}
{\Large \bf 
Neutrino degeneracy effect on neutrino oscillations\\ 
and primordial helium yield}\\

\vspace{.5cm}
{\large D. P. Kirilova}\footnote{Permanent address:
 Institute of Astronomy, Bulgarian Academy of Sciences,\\
blvd. Tsarigradsko Shosse 72, Sofia, Bulgaria\\
E-mail:$mih@phys.uni$-$sofia.bg$,
$dani@libra.astro.acad.bg$}

\vspace{.5cm}

{\em The Abdus Salam\\
International Center for Theoretical Physics\\
 Strada Costiera 11, 34014  Trieste, Italy
}

\vspace{1.5cm}

{\large M. V. Chizhov}\footnote{Permanent address:
Centre for Space Research and Technologies, Faculty of Physics,\\
University of Sofia, 1164 Sofia, Bulgaria\\   
E-mail: $mih@phys.uni$-$sofia.bg$}

\vspace{.5cm}

{\em The Abdus Salam\\
International Center for Theoretical Physics\\
 Strada Costiera 11, 34014  Trieste, Italy
}
\end{center}
\vfill

\begin{abstract}

In this work we present the results of our numerical analysis
of the effect of lepton asymmetries $L$ on 
neutrino oscillations and consequently on helium production in a model of
cosmological nucleosynthesis with  active-sterile
neutrino oscillations.
We performed detail calculations of the neutron-proton ratio till its 
freezing and the subsequent synthesis of helium-4  for a wide range of 
values of the initial lepton asymmetry $10^{-4}-10^{-10}$ and for 
the complete set of mixing parameters of the oscillation model described 
in \cite{our}: $\delta m^2\le 10^{-7}$ eV$^2$ and any $\vartheta$. 

We have found that there are significant modifications of neutrino 
number densities and energy distributions over a large range of values of 
the initial asymmetry. Hence, the previously derived 
(for the case of lepton asymmetry of the order of the baryon one)  
nucleosynthesis limits on the oscillation parameters are  
considerably altered. 
We have calculated the dependence of primordially produced helium-4 on 
the parameters of the model $Y_p(L, \delta m^2,\vartheta)$. From the  
isohelium contours in the $\delta m^2-\vartheta$
plane, we have obtained limits on the neutrino mixing parameters
corresponding to nucleosynthesis with degenerate neutrinos.
An intriguing new result
of our analysis is that the lepton asymmetry is able to  {\it tighten} the 
nucleosynthesis bounds on neutrino mixing parameters, besides its well 
known ability to relax them.  

\end{abstract}

\ \\

PACS number(s): 05.70.Ln, 14.60.Pq, 26.35.+c

keywords: neutrino oscillations, lepton asymmetry, cosmological nucleosynthesis

\vfill

\newpage

\twocolumn

\section{Introduction}
 Lepton asymmetry is not directly observable. Besides,  
in general, large lepton numbers L bigger than 
the baryon number B ($L > B$) and even $L \gg B$ 
are not excluded  by any profound theoretical principle. 

The origin of eventual 
big lepton numbers is not fully understood.  
There exist numerous different mechanisms for their generation. 
Large lepton numbers can be consistent with the observed small baryon number 
for example in the context of Grand unification, as pointed in 
\cite{origin1}; there exist different models producing $L \gg B$: in a  
scenario of baryogenesis with baryonic charge condensate~\cite{dkbaryon},
where large and varying lepton asymmetry can be produced,
 the out-of equilibrium decays of heavy Majorana neutrinos
\cite{fuya,mu} can produce it as well, 
resonant neutrino oscillations can create considerable neutrino asymmetry
at low temperature~\cite{create,our}. 
See also the review papers~\cite{drev,sarkar}. 

Besides, large lepton asymmetry has interesting implications 
in the early Universe: in baryogenesis models
~\cite{linde,fuya,baryon}, for solving 
the monopole problem \cite{monopol},
 as well as domain wall problem \cite{domain},
in primordial nucleosynthesis~\cite{lnuc1,lnuc2,nucomega}
for structure formation in the Universe~\cite{structure},
relic asymmetry can suppress neutrino transitions in the 
early Universe and have interesting implications for 
solar neutrino problem, as well as in cosmological nucleosynthesis (CN)
~\cite{suppress}.

Therefore, in this work we consider it worthwhile  to explore 
precisely the effect of  lepton 
asymmetries and to study      
the problem of oscillations and their influence 
on primordial production of helium-4 in the presence of such  
asymmetries. Especially, in 
 the present work we would like to show  that much smaller 
asymmetries ($|L| \ll 0.1$) still can considerably  
affect cosmological nucleosynthesis, although indirectly through its 
effect on neutrino oscillations. Thus, the presence of an initial 
asymmetry bigger than the baryon one may considerably change the bounds  
on oscillation parameters obtained  from primordial nucleosynthesis 
considerations.

The theme of lepton asymmetry in nucleosynthesis models with oscillations 
was previously considered in several publications
~\cite{bd2,kimmo,savage,suppress,our,our2,create}. 
The possibility of changing the 
nucleosynthesis constraints on lepton asymmetry due to oscillations was 
discussed within models with great $L \sim 0.1$~\cite{savage}. The dynamical 
evolution of $L$
due to oscillations was studied for some particular values of $L$ and 
neutrino mixing parameters~\cite{bd2,kimmo}. The possibility for 
loosening the nucleosynthesis bounds on mixing parameters due to the 
suppression of oscillations by great enough $L$ was 
revealed~\cite{suppress}. And also the possibility of generation of 
asymmetry due to resonant oscillations was  studied
~\cite{our,create}, and some scenarios where oscillations generate 
asymmetry which further suppresses oscillations and as 
a result alleviates nucleosynthesis bounds on neutrino mixing  were
discussed.
However, always the simplifying assumption that the evolution of the 
neutrino ensemble follows the average neutrino momentum was made.
And the case of high temperature resonant oscillations $\delta m^2 < 0$ 
with big mass differences  and great asymmetries 
  was considered. The nonresonant case was considered just for 
a certain set of parameters and for not very large initial $L$ it was 
shown that neutrino asymmetry adjusts dynamically to zero~\cite{kimmo}.  

 In this work we consider the  case of 
nonresonant oscillations 
$\delta m^2 > 0$ with small mass differences for a wide range of the 
initial values of $L$, namely $10^{-10}-1$, and provide a precise 
numerical study for the most unexplored part of it: $10^{-7} 
\le |L| \le 10^{-4}$. 
We insist here for the correct account for the spectral spread of 
neutrino, 
which we  find to play an important role in the oscillation-asymmetry 
interplay, and to be decisive  for the realization of 
low temperature resonance in $\delta m^2 > 0$ case.\footnote{Mind that, 
as was shown in the pioneer work~\cite{kimmo}, in models 
working with  the mean momentum only,  the resonance does not have 
place as far as actually the asymmetry calculated in this approximation 
reaches zero at resonance.}
 
We find that even 
small asymmetries $|L| > 10^{-7}$ may considerably effect nucleosynthesis 
(contrary to the conclusions of the previous works) 
and that asymmetry is also able 
to {\it enhance oscillations and consequently to strengthen}
 the nucleosynthesis bounds on neutrino mixing parameters 
for  concrete regions
in the parameter space ($L, \delta m^2,\vartheta$),
besides suppressing oscillations and thus loosening the
bounds on mixing parameters from nucleosynthesis.
The analysis of the asymmetry effect is provided using the precise kinetic 
approach to the problem of neutrino oscillations and asymmetry, described 
in~\cite{our,our2}. Actually, the present work is an extension of our 
previous study ~\cite{our,our2} for the
case of a lepton asymmetry bigger than the baryon one.
 There we have calculated the bounds on the neutrino mixing parameters
for a $\nu_e \leftrightarrow \nu_s$ mixing scheme with small mass
differences $\delta m^2 \le 10^{-7}$ eV$^2$, where $\nu_e$ is the active
electron neutrino, while $\nu_s$ denotes the sterile antineutrino.
This case corresponds to
nonequilibrium oscillations between electron
neutrinos $\nu_e$ and sterile neutrinos $\nu_s$
when the latter do not thermalize till $\nu_e$ decoupling at
2 MeV and oscillations become effective after $\nu_e$ decoupling.  
We  discussed the special case of {\it nonequilibrium oscillations}
between weak interacting electron neutrinos and sterile neutrinos for
small mass differences $\delta m^2$, as far as the case of large 
$\delta m^2$ was already studied (both concerning small asymmetries 
$L\sim B$ ~\cite{bd1,ekt,ekm,cl,ssf,s} 
and bigger ones $L > B$~\cite{kimmo,suppress,create}).
 Due to the correct kinetic approach accounting for
neutrino depletion, neutrino spectrum distortion and neutrino asymmetry
for each neutrino momentum, we have obtained more precise
 bounds (an order of magnitude stronger than the existing previously)
 to the neutrino oscillation parameters. In
the present work we will show how  these bounds
are  changed due to bigger lepton asymmetries.
           
We would like to emphasize that the qualitatively 
new result, namely that small asymmetries may enhance 
oscillations and consequently to strengthen the nucleosynthesis bounds, 
was revealed only due to the correct approach accounting for 
the spectral spread of neutrino momenta and energy spectrum distortion 
due to oscillations, advocated in our previous papers~\cite{our,our2,dpk}
and in the pioneer work of Dolgov~\cite{do}.
Other studies missed this possibility as far as there it was assumed 
that neutrino ensemble follows the behavior of average momentum
even in the presence of oscillations and asymmetry. 

In this work  we explore  the different ways in which lepton asymmetry 
 may affect oscillations (and vice versa - oscillations affect the 
asymmetry) and primordial nucleosynthesis. 
We have defined numerically the  concrete $L$ ranges where these 
effects hold, namely: 
(a) big enough L -- total suppression of oscillations
corresponding to degenerate nucleosynthesis, without oscillations 
(b) resonance region -- L enhancing oscillations and leading to an 
overproduction of He-4, and  (c) small L -- negligible effect.

The paper is organized as follows. In the following 
Section II we  discuss  neutrino degeneracy.
In Section III we present the model of nonequilibrium neutrino
oscillations and analyze the neutrino evolution, using  kinetic 
equations for the neutrino density matrix for each momentum mode.
In Section IV we investigate $\nu_e \leftrightarrow \nu_s$
the primordial production of helium  in the presence of 
oscillations and neutrino degeneracy.
The results and conclusions are presented in Section V.

\section{Lepton asymmetry - Preliminaries}

At the epoch of interest - just prior to cosmological nucleosynthesis
- at temperatures around few MeV,
the lepton asymmetry is expressed through the asymmetries in the 
neutrino sector and the electron sector:
$$
L=\sum_\ell(N_\ell-N_{\bar{\ell}})/N_{\gamma},
$$
where $\ell=e, \nu_e, \nu_{\mu}, \nu_{\tau}$, 
and $\bar{\ell}$ corresponds to their antiparticles.
Big lepton asymmetry $L > B$ may be contained only in the
neutrino sector,  
as far as the electron-positron asymmetry must be equal to the 
proton-antiproton asymmetry from the charge conservation requirements.
Hence, it will be more correct to talk about 
neutrino asymmetry or neutrino degeneracy, as it is most often  
referred to. 
It is convenient instead of $(N_{\nu}-N_{\bar{\nu}})/N_{\gamma}$
to use the so called neutrino degeneracy parameter, $\xi=
\mu/T$, where $\mu$ is the corresponding neutrino chemical 
potential. 
In the thermal equilibrium of the early Universe at $T=2$ MeV,
the neutrino degeneracy for the case 
$\mu/T \ll 1$ is given by 
$$
L=\sum_i (N_{\nu_i}-N_{\bar{\nu}_i})/N_{\gamma}\sim
\sum_i (\pi^2/12 \zeta(3))(T_{\nu_i}/T_{\gamma})^3\xi_i
$$
where $i= e, \mu, \tau$.

There are no experimental or direct theoretical 
limitations for the magnitude or the sign of neutrino asymmetries.
However, there exist indirect limits on the neutrino 
degeneracies in the different neutrino flavors, 
obtained on the basis of astrophysical and cosmological 
considerations. Such are the constraints  obtained from 
the present age and density of the Universe~\cite{age}:  
 $\xi \le 86$
for the degeneracy in only one neutrino species;
from the requirement for long enough matter dominated period, 
necessary for successful structure formation:
 $\xi  \le 6.9$, 
 and the strongest bound on the magnitude of L is  obtained from 
 primordial nucleosynthesis considerations~\cite{lnuc1,lnuc2,ho}: 
$-0.06\le \xi_e \le 1.1$ and $|\xi_{\mu}| \le 6.9$ 
for baryon to photon ratio $\eta \le 1.9 \times 10 ^{-10}$.

As far as primordial nucleosynthesis is 
particularly sensitive both to the neutrino degeneracies and to neutrino 
oscillations, it is interesting to consider the effect of 
neutrino degeneracies in a modified model of primordial nucleosynthesis 
{\it with neutrino oscillations}. 

In this work we  assume no degeneracy of muon and tau
 neutrinos for simplicity. However, this assumption is not essential.
The results can be easily rescaled for the general case of asymmetry in
all sectors.
We  consider the effect of neutrino asymmetries in the wide diapason 
of its initial values $10^{-10}-1$. Both positive and negative values of 
$L$ are considered.
 Asymmetries smaller than $10^{-10}$ have negligible 
effect, therefore we have not studied them. 
Now we have found that small asymmetries with magnitude
$|L| < 10^{-7}$ have negligible effect\footnote{This confirmed 
our preliminary results stated in ~\cite{our2}.} 
also for the concrete mixing parameters of our model.
We have proved that asymmetries
 greater than $10^{-4}$ completely suppress the oscillations and
affect the nucleosynthesis due to the pure direct effects of neutrino
degeneracy on nucleosynthesis - through affecting
the expansion rate and the weak reaction rates. The latter effects have been 
studied already in the numerous publications on degeneracy and
nucleosynthesis~\cite{lnuc1,lnuc2,ho}, therefore,  we have not provided 
detail numerical calculations for such big asymmetries.
 While asymmetries in the intermediate 
region, namely $10^{-7} \le |L| < 10^{-4}$ have not been studied 
systematically. 

In this work we prove that such asymmetries can both 
enhance and suppress 
oscillations depending on the concrete set of mixing parameters. For that 
range we 
have provided detail numerical analysis of the effect of the neutrino 
degeneracy  on neutrino evolution and consequently on 
helium-4 production. 
Finally (as described in the following sections) we have obtained 
the isohelium contours for different $L$ values in the $(\delta m^2,
\vartheta)$
plane and derived  new limits on the neutrino mixing parameters for 
the cases with neutrino degeneracy.

 \section {Evolution of the neutrino ensembles in the 
presence of initial neutrino degeneracy}

\subsection{Neutrino oscillations - the model}

We suppose the existence
of a sterile neutrino ($SU(2)$-singlet)  $\nu_s$, and would 
like to explore the
cosmological effect
of {\it nonresonant neutrino oscillations} $\nu_e \leftrightarrow
\nu_s$ on the primordial nucleosynthesis
for a wide range of values of the initial lepton asymmetry. 
 
Within the model of interest oscillations proceed effectively after 
the active neutrino decoupling and till then the sterile neutrinos 
have not yet thermalized\footnote{This assumption of nonthermalization 
of steriles till electron neutrino decoupling and the late effectiveness of
oscillations limits the
allowed range of oscillation parameters for the discussed  model:
 $\sin^2(2\vartheta) \delta m^2 \le 10^{-7}$ eV$^2$ \cite{our}.
However, it allowed us to examine the interesting nonequilibrium case of 
neutrino
oscillations,in general with non Fermi-Dirac distribution,  which is 
realized at such  small mass differences.  And besides,
it allowed us to neglect the second order in $G_F$ terms in the equations 
governing the neutrino evolution, and helps simplifying the calculation 
process.
(For the nonequilibrium case studied, this is important, as far as
accounting for the neutrino spectrum distortion for that case is 
obligatory, and it needs much longer calculation time.)}
so that their number density is negligible in comparison with the 
electron neutrino one.
The model is described in detail in~\cite{our,our2}.
 
Oscillations between  $\nu_s$
($\nu_s \equiv \tilde{\nu}_L$) and the active   
 neutrinos proceed according to the Majorana\&Dirac ($M\&D$)
mixing scheme~\cite{b}. For simplicity  mixing present just in the electron 
sector is assumed $\nu_i=U_{il}~\nu_l$, $l=e,s$:
\begin{eqnarray*}
\nu_1 & = & c\nu_e+s\nu_s,\\
\nu_2 & = & -s\nu_e+c\nu_s,
\end{eqnarray*}
where $\nu_s$ denotes the sterile electron antineutrino,
$c=\cos(\vartheta)$,
$s=\sin(\vartheta)$ and $\vartheta$ is the mixing angle in the electron 
sector, the mass eigenstates $\nu_1$ and  $\nu_2$ are
Majorana particles with masses correspondingly $m_1$ and $m_2$.
We consider the nonresonant  case $\delta m^2=m_2^2-m_1^2 > 0$,
which corresponds in the small mixing angle limit to a sterile neutrino
heavier than the active one. We would like to remind, that `nonresonant' 
is used in the sense that it does not allow resonant transitions 
due to the finite temperature correction term (nonlocal term). 
 However, as it is obvious, for the low temperature resonance 
 this terminology 
is misleading. As far as for  any sign of $\delta m^2$ then there is a 
possibility for a resonance due to the local term either in the neutrino 
sector or in the antineutrino sector.
 
 \subsection{The kinetics of nonequilibrium neutrino oscillations}

 The kinetic equations for the density matrix of the nonequilibrium
oscillating neutrinos in the primeval plasma of the Universe
in the epoch previous to nucleosynthesis, i.e. consisting of
photons, neutrinos, electrons, nucleons,
and the corresponding antiparticles,
have the form:
\be
{\partial \varrho(t) \over \partial t} =
H p~ {\partial \varrho(t) \over \partial p}
+ i \left[ {\cal H}_o, \varrho(t) \right]
+i \left[ {\cal H}_{int}, \varrho(t) \right],
\label{kin}
\ee
where $p$ is the momentum of electron neutrino and $\varrho$ is the
density matrix of the massive Majorana neutrinos in momentum space.

These equations                                     
account   {\it simultaneously} for the participation of neutrinos
into expansion, oscillations and interactions with the medium.    
The first term in the equation describes the effect of expansion,
the second is responsible for oscillations, the
third accounts for forward neutrino scattering off the
medium. The collisions (second order interactions in $G_F$
of neutrinos with the medium) are neglected, as far as we consider 
the case of oscillations effective after neutrino decoupling.

 ${\cal H}_o$ is the free neutrino Hamiltonian in the eigenstate basis: 
$$ 
{\cal H}_o = \left( \begin{array}{cc}
\sqrt{p^2+m_1^2} & 0 \\ 0 & \sqrt{p^2+m_2^2}
\end{array} \right),
$$
while ${\cal H}_{int} = \alpha~V$ is the interaction Hamiltonian,
where $\alpha_{ij}=U^*_{ie} U_{je}$,
$V=\sqrt{2} G_F \left(+{\cal L} - Q/M_W^2 \right)N_\gamma$.

The first `local' term in $V$ is proportional to
the fermion asymmetry of the plasma and for $L > B$ is 
essentially expressed through the neutrino asymmetry:
$$
{\cal L} \sim 2 L_{\nu_e}+ L_{\nu_{\mu}} + L_{\nu_{\tau}}.
$$

 The second `nonlocal' term is momentum dependent 
$Q \sim E_\nu~T$ 
~\cite{nr,bd1}.  When lepton asymmetries of 
the order of the baryon one are considered, the nonlocal 
term dominates at high temperature, while with cooling of the Universe in 
the process of expansion the local one becomes more 
important~\cite{bd1,kimmo,bd2}.
 For greater asymmetries $L > B$ the role of the nonlocal term 
decreases in comparison with the local one,
and usually for temperatures less than a few MeV it is neglected. 
We have precisely accounted for it in this work, having in mind the following.
Due to the momentum dependence of the nonlocal 
term  even when for the mean  neutrino momentum modes 
at a given temperature this term is smaller than the local one, for 
the high energetic modes it may  still play some role.   
We have checked numerically that the nonlocal term can change from a 
few up to about  14\% 
the results on helium-4 production. The maximum effect being, as can 
be easily guessed, in the case of maximum mixing. This is noticeable.
 
An analogues equations hold for the antineutrino density matrix,
the only difference being in the sign of the fermion asymmetry: 
${\cal L}$ is replaced by $-{\cal L}$.

Medium terms depend on neutrino energy density and neutrino degeneracy,
thus introducing a nonlinear feedback mechanism.
As was already observed in previous works
~\cite{suppress,our,kimmo}, oscillations change
neutrino-antineutrino asymmetry and it in turn affects
oscillations. For large $L$
the evolution of neutrino and antineutrino ensembles is
strongly coupled and hence, it must be considered simultaneously.

The equation (\ref{kin}) results into a set of coupled nonlinear
integro-differential equations with time dependent coefficients
for the components
of the density matrix of neutrino. Due to the greater $L$ the coupling is 
stronger than for the  case $L=10^{-10}$. 
As far as for these strongly 
coupled nonlinear equations an analytic solution is hardly possible, we 
have provided an exact 
kinetic analysis of the neutrino evolution by a numerical integration of
the {\it kinetic equations
  for the neutrino density matrix for each momentum mode}.
The evolution
of the neutrino ensembles is followed numerically from the $\nu_e$ 
freezing at 2 MeV till the formation of helium-4. 

Our numerical analysis shows that the coupling between the neutrino and 
antineutrino ensembles  leads to their similar behavior:  
Whenever the resonance condition is fulfilled for neutrino 
(or antineutrino), and the ensemble suffers a 
resonant oscillations, due to the strong coupling  
between the systems, the antineutrino ensemble  
shows the same behavior (after some negligible delay time) 
too.\footnote{Opposite to the traditional naive conclusions,
based on observations just for a certain parameter values
and provided for the mean neutrino momentum, that in the low 
temperature resonance case (when the nonlocal term is neglected in 
comparison with the local one) the resonance may take place either 
in the neutrino sector or in the antineutrino one, but not 
both~\cite{kimmo}.}
The results have almost negligible dependence on the sign of the 
initial asymmetry, as could be expected from this behavior of the 
ensembles. 
    The concrete `spectral' mechanism of that  strong coupling between the 
ensembles  in our case is described in detail further on. 

The initial condition for the neutrino ensembles in the interaction basis
is assumed of the form:
$$
\varrho = n_{\nu}^{eq}
\left( \begin{array}{cc}
1 & 0 \\
0 & 0
\end{array} \right)
$$
where $n_{\nu}^{eq}=\exp(\xi-E_{\nu}/T)/(1+\exp(\xi-E_{\nu}/T))$.

It corresponds to an  equilibrium distribution of 
active electron neutrinos, and an absence of the sterile one. 
 We have analyzed the evolution of nonequilibrium oscillating
neutrinos by numerically integrating the kinetic equations 
 for the period after
the electron neutrino decoupling till the freeze out of the
neutron-proton ratio
($n/p$-ratio), i.e. for the temperature interval 
 from 2 MeV till 0.3 MeV.
We have explored the problem using the Simpson method for
integration and the  fourth order
Runge-Kutta algorithm for the solution of the differential
equations. 
In order to keep the appropriate 
precision of calculation for the wide range of initial
parameter values, we used a step for the evolution with  temperature  
decrease of the order $10^{-7}-10^{-6}$ MeV and for the calculation of 
the neutrino spectrum a step for $E_\nu/T$ of the order $10^{-2}-10^{-3}$.

The oscillation parameters range
studied is $\delta m^2 \in [10^{-11}, 10^{-7}]$ eV$^2$
 and $\vartheta \in[0,\pi/4]$, which is the full set of mixing 
parameters in the discussed model ~\cite{our2} of oscillations. 
The neutrino degeneracy was varied for a wide  range $10^{-10}-10^{-4}$
of $L_{\nu}$ as discussed in the previous chapter.

The distributions of electrons and positrons were taken the
equilibrium ones.  
The neutron and proton number
densities, used in the kinetic equations for neutrinos, were substituted
from the numerical calculations in CN code accounting for
neutrino oscillations. I.e. we have simultaneously solved the
equations governing the evolution of neutrino ensembles and those
describing the evolution of the nucleons (see the next section).
The baryon asymmetry $\beta$, parameterized as the ratio of the baryon number
density to the photon number density,
was taken to be $3\times 10^{-10}$.
 
\section{Nucleosynthesis with nonequilibrium oscillating neutrinos}

We have calculated the production of helium-4 in a detail model of primordial
nucleosynthesis, accounting for the direct kinetic effects of oscillations
and neutrino degeneracy on the neutron-to-proton transitions.
The effect of oscillations on helium-4 has been discussed in numerous 
publications~\cite{hp,la,MSW,bd2,kimmo,savage,our2,bd1,ekt,ekm,cl,ssf,s,dpk}.
We follow the approach described in~\cite{our,our2}.
The numerical analysis is performed working with exact kinetic
equations for the nucleon number densities and neutrino density matrix
in momentum space. This enabled us to analyze the direct influence of
oscillations onto the kinetics of the neutron-to-proton transfers
and to account precisely for the 
neutrino population depletion, distortion of the neutrino spectrum and the
role and dynamical evolution  of neutrino-antineutrino asymmetry.

The master equation, describing the evolution of the neutron
 number density in momentum space $n_n$ for the case of oscillating
neutrinos $\nu_e \leftrightarrow \nu_s$, reads:
\begin{eqnarray}
&&\left(\partial n_n / \partial t \right)
 = H p_n~ \left(\partial n_n / \partial p_n \right) +
\nonumber\\
&& + \int {\rm d}\Omega(e^-,p,\nu) |{\cal A}(e^- p\to\nu n)|^2
\nonumber\\
&&~~~~~~~ \times\left[
n_{e^-} n_p (1-\varrho_{ee}) - n_n \varrho_{ee} (1-n_{e^-})\right]   
\nonumber\\
&& - \int {\rm d}\Omega(e^+,p,\tilde{\nu}) |{\cal A}(e^+n\to p\tilde{\nu})|^2
\nonumber\\
&&~~~~~~~\times\left[
n_{e^+} n_n (1-\bar{\varrho}_{ee}) - n_p \bar{\varrho}_{ee}
(1-n_{e^+})\right].
\end{eqnarray}

We have calculated the neutron-to-photon ratio evolution prior to 
nucleosynthesis epoch, namely 
  for the temperature interval 
  $T \in [0.3,2.0]$ MeV and for the full range of oscillation
parameters of our model and $L$ in the range $10^{-10}-10^{-4}$.  The 
value of $\varrho_{ee}(E_\nu/T)$ at each integration step was taken 
from the simultaneously performed integration of the
set of equations for neutrinos, i.e. the evolution of neutrino and the
nucleons was followed self consistently.

The initial values at $T=2$ MeV for the neutron, proton and electron
number densities are their equilibrium values.The parameters values of 
the CN model, adopted
in our calculations,  are the following: the mean neutron lifetime is
$\tau= 887$ sec, which corresponds to the present weighted average
value, the effective number of relativistic flavor types of
 neutrinos
during the nucleosynthesis epoch $N_{\nu}$ is assumed equal to the
standard value $3$.

On the basis of this analysis the primordially 
produced He-4 value was obtained as a function of 
 neutrino degeneracy parameter and 
neutrino mixing parameters. 

\section{Results and conclusions}

We have calculated the evolution of the neutron-to-nucleon number densities
 $X_n(t)=N_n(t)/(N_p+N_n)$ 
for each set of values in ($L,\delta m^2,\vartheta$) parameter space.
In Figs.~1 the evolution of 
$X_n$ is illustrated for initial asymmetry of the order of the baryon 
one and for $L=10^{-6}$ and $L=10^{-5}$.
As is seen from the figures, even relatively small asymmetry ($L=10^{-6}$)
can considerably effect the neutron-to nucleon ratio.
In some cases, i.e. for maximal mixing  (Fig.~1a) leading to
underproduction of helium-4, while in others (Fig.~1b) to
its overproduction in comparison with helium yields calculated
in models with small asymmetry ($L\sim 10^{-10}$).

In Fig.~2 the dependence of the frozen neutron
number density relative to nucleons $X_n^F(L)$ on the
the initial neutrino asymmetry for $\sin^2(2\vartheta)=10^{-0.05}$
and different fixed  $\delta m^2$ is illustrated.
The dependence of the frozen neutron
number density relative to nucleons $X_n^F(\vartheta)$ on the
mixing angle for different fixed $\delta m^2$, is presented in Fig.~3.

It is interesting to note, that, as illustrated in Fig.~3
depending on the mixing angle, it is possible to weaken or
to strengthen  the limits in comparison with the
model of nucleosynthesis with oscillations without a considerable
lepton asymmetry. Namely,
for large mixing, the presence of neutrino degeneracy of the order
$10^{-6}$, for example, leads to a less strong constraints on 
the mixing parameters, weakening the oscillation effect of
overproducing helium. While for smaller mixing angles the same asymmetry
leads to stronger limits on mixing parameters, as a result of the
enhancement of oscillations, and hence increasing the helium
overproduction.

These results can be understood as follows:
As it was shown in ~\cite{our2} in case of baryon like small 
lepton asymmetry the kinetic effects
(neutrino population depletion and distortion of neutrino spectrum)
due to oscillations play an important
role and lead to a considerable overproduction of helium.

In the presence of a bigger asymmetry both the neutrino depletion and 
spectral distortion are changed due to the interconnections between 
oscillations and asymmetry.
Large enough asymmetries (in our case 
$L > 10^{-7}$) affect oscillations, while strong enough oscillations 
also influence asymmetry and lead to its dynamical evolution. 
We would like to note, that even for relatively small $|L| < 10^{-5}$
the effect of asymmetry is considerable (contrary to the thought 
before). And also that even oscillations with small mass differences 
are able to influence neutrino asymmetry.
A characteristic situation in 
our model is that great asymmetries cannot be created: 
neutrino asymmetries typically of an order of magnitude greater than the 
initial one are generated, however they are rapidly oscillating, the 
average value being zero. In our case of nonresonant oscillations,
the presence of the second system (for antineutrino) hinders the 
continuous growth of asymmetry.
In more detail asymmetry-oscillation relations 
will be analyzed in a following paper.
 Our main purpose here was  
the study of asymmetry effect on primordial production of helium-4, 
and derivation of  the nucleosynthesis limits on neutrino mixing in the 
presence of neutrino degeneracy. 

For the oscillation model discussed we have found three
interesting regions for the range of the lepton asymmetry.
To be more clear let us consider nearly  maximal mixing: 
$\sin^2(2\vartheta)=10^{-0.05}$ (see Fig.~2 for illustration).
 Then these  regions are as follows.

A. In the range $|L_{\nu}|
\ge 5 \times 10^{-5}$ the asymmetry fully suppresses neutrino  
oscillations.\footnote{This result is in accordance with the
estimations made in ~\cite{suppress},
 where it was shown that
a big neutrino asymmetry may lead  to suppression
of oscillations.}
The yield of helium-4 coincides with the values
obtained
in the models of primordial nucleosynthesis without oscillations.
Therefore, the nucleosynthesis bounds on the mixing parameters
in the presence of such large asymmetries are 
waved away.  

For $|L| \le 10^{-2}$ the direct kinetic effect of such asymmetries on the   
neutron-to-proton transfer is almost negligible, i.e.
for the range $10^{-4}-10^{-2}$ the only role of asymmetry is to 
suppress oscillations.
Asymmetries greater than that effect nucleosynthesis by directly 
changing the kinetics of nucleons and are exhaustively studied
in previous works~\cite{lnuc1,lnuc2,ho}.
And the effect of the asymmetry is well known:
Neutrino degeneracy effects element
production in two ways~\cite{lnuc1}.
 First, the nonzero chemical potential results to an
increase of neutrino energy density and hence to an increase of the
expansion rate of the Universe thus allowing less time for the nuclear
reactions to proceed. Second, it alters the reaction rates governing
the neutron-proton transitions, as far as these rates are extremely
sensitive to the neutrino spectrum. Namely, electron neutrino degeneracy
results into preponderance of protons, hence into a drop of
primordial helium-4 abundance, while electron antineutrino degeneracy
leads to the increase of neutrons.
Therefore for $L > 10^{-4}$ we have not provided detail numerical 
calculations.

B. The range of smaller than $10^{-4}$ asymmetries was more 
appealing to us 
as far as it was totally unexplored till now. Therefore,   
we  numerically analyzed the problem for the initial
values of the neutrino asymmetry in the range $10^{-10}-10^{-4}$,
and for the full mixing parameter space of the active-sterile
oscillations model, described in~\cite{our}, i.e.
for all mass differences $\delta m^2 \le 10^{-7}$ eV$^2$ and   
 mixing angles $\vartheta$.

The  zone $10^{-7} \le |L| \le 5 \times 10^{-6}$
is the most interesting one. The asymmetry with magnitudes of that order
is strong enough to influence neutrino oscillations.
And although the
neutrino asymmetry is too small to effect nucleosynthesis directly,
it does effect nucleosynthesis yields in indirect way
 through its influence on neutrino oscillations pattern.
The asymmetry values in that range, depending on the concrete values
of mixing parameters, are able  to enhance oscillations. 
So, depending on the mixing angle for some L in this
interval the ensemble of neutrinos expires a resonance.
For nearly maximal mixing it is roughly around $L \sim 10^{-6}$ (Fig.~2).
This enhancement of oscillations
is  big enough to influence considerably the
electron neutrino and electron antineutrino number density and spectrum 
i.e. it leads to an enhanced depletion of the number densities of neutrinos 
and antineutrinos(!) and  
a decrease in their mean energy, as well as nontrivial evolution 
of the asymmetry 
itself. Hence influencing the helium-4 primordial value, which is extremely 
sensitive to it. The result is 
overproduction of helium-4 in comparison with the  case of primordial
nucleosynthesis in the presence of oscillations with negligibly small
L. This on its turn leads to a strengthening of the nucleosynthesis
bounds on the neutrino mixing parameters for that certain set of 
parameters ($L, \delta m^2, \vartheta$).

The ability of $L$ within that range to  enhance  the oscillations  
looks quite amazing at a first glance. The naive picture one expects  as 
a result of increasing the initial $L$ is  a gradual suppression of
oscillations proportional to $L$. As far as  $L$ value, calculated for  
the  mean neutrino momentum is by orders of magnitude bigger than 
the necessary one for a resonance transfer $L \gg L_r$. 

However, the detail numerical analysis, accounting for the momentum 
spread of neutrino ensemble, showed a more complex picture. Varying $L$ 
for fixed mixing parameters we have observed a resonance region, i.e. an 
enhancement of oscillations, as seen in Fig.~2 and Fig.~3.
And besides that, when a resonance for a given $L$ value was observed  
for neutrino it was followed with a negligible time delay by a 
resonance in antineutrino! This leads to a considerable depletion of 
number densities and to further overproduction of helium-4.
This amazing results can be explained as follows.   

The system of nonlinear differential equations cannot be solved 
analytically without radical assumptions. However, the qualitative 
behavior of the ensembles and the obtained results concerning helium-4 
can be guessed from the following simplified considerations.

The resonant conditions for the neutrino looks like~\cite{MSW}:
$$
\cos(2 \vartheta) (\delta m^2/2 E_{\nu})= \sqrt{2} G_F(L-Q/M_W^2) N_{\gamma}
$$
while for antineutrino the sign before $L$ term is the opposite one. 

For the mean neutrino momenta $\bar{p} = 3.15~T$ 
in our model
\begin{eqnarray*}
&|L|& \gg Q/M_W^2\\ 
&|L|& \gg \cos(2 \vartheta) \delta m^2/ (2\sqrt{2} E_{\nu} G_F N_\gamma).
\end{eqnarray*}

{\noindent}Now let us consider the possibility that at a given temperature, 
for neutrino of a given momentum $p < \bar{p}$ the resonance condition is 
fulfilled, i.e. $L$ has the resonance value $L_{r}(p)$. 
Then neutrinos with this momentum suffer a resonant transfer, leading 
to a decrease in the number densities of neutrinos (in favour of 
the sterile neutrinos). 
As far as for antineutrinos $L$ has the opposite sign, oscillations 
remain suppressed, and the number densities of antineutrinos 
do not change.
Hence, the net result is a decrease in $L$ due to this resonant transition. 
This decrease makes possible the 
fulfillment of the resonant condition for more energetic neutrinos, 
leading to further decrease of $L$ and so on;
due to this `resonance wave' passing to neutrinos with higher and higher 
momenta, the neutrino number densities expire a considerable depletion,  
consequently, having in mind the small initial values of the neutrino 
asymmetry ($\delta N_{\nu} \sim L_{\nu} N_{\nu}$) soon this resonant 
wave leads to a change in the sign of $L$. This suppresses 
further resonant transfer for neutrinos, however, for the 
antineutrino ensemble now there appears the possibility for a resonance.
So, the same process follows in the antineutrino ensemble.     
The only difference being that contrary to the neutrino system, for 
antineutrinos the resonance is first fulfilled for the high  energy 
neutrinos and then passes to the low energetic ones - i.e. 
it reaches more rapidly neutrinos with mean momentum, and the process is 
more avalanche like. 
 This rapid decrease of antineutrinos leads to a rapid 
bump of $L$ which again becomes positive. This pendulum like process
proceeds relatively fast, the oscillation of the asymmetry proceeds much 
faster than the temperature decrease. 
The total effect is that the resonant transfer both for neutrino and 
antineutrino system is realized even for a considerable initial 
asymmetry values (nonresonant ones, when calculated for the mean 
neutrino momentum), and due to the enhanced transfer of active to sterile 
neutrinos a considerable `resonant' production of helium-4 is observed. 

C. For small asymmetries $|L| < 10^{-7}$ we have confirmed our preliminary
results from ~\cite{our2}, namely that asymmetry with these values has 
too week effect on nucleosynthesis to be considered.

Finally, we have calculated the He-4 dependence on
the neutrino asymmetry for the whole range of mixing parameters of 
our model and we have obtained the primordial helium yields
$Y_p(L,\delta m^2,\vartheta)$.
 This enabled us to obtain isohelium contours in the $\delta m^2-\vartheta$
plane. Some of these  constant helium
contours, for $L=10^{-10}$ and $L=10^{-6}$, are presented in Fig.~4.

 On the basis of these results, requiring an agreement between the
 theoretically predicted and the inferred from observations 
 values of primordial helium,  
  we have  obtained  cosmological
constraints on the neutrino mixing parameters corresponding to 
different initial lepton asymmetry values.
Assuming the conventional observational bound on primordial $^4\! He$
$Y_p=0.24$
 the cosmologically excluded  region for the oscillation parameters
on the plane $\sin^2(2\vartheta) - \delta m^2$ in Fig.~4
 lies to the right of  the $Y_p=0.245$ curve, which
gives $5\%$  overproduction of helium in comparison with the
standard value. For comparison the curves corresponding to $L=10^{-10}$ 
and $L=10^{-6}$ are plotted.

In conclusion we would like to stress once again that 
lepton asymmetries $|L|\ge 10^{-7}$ 
can both enhance or suppress (completely or partially, depending on the
concrete values of the model parameters) 
 neutrino oscillations and consequently can strengthen 
or weaken oscillation effect on primordial nucleosynthesis. 
Therefore,  the constraints on the oscillation parameters 
obtained from nucleosynthesis  with  lepton asymmetry of the order of 
the baryon one~\cite{bd2,kimmo,savage,our2,bd1,ekt,ekm,cl,ssf,our}
can be considerably changed - 
either relaxed (or even totally removed) or strengthened for 
(note!) {\it not very large} values of the initial lepton asymmetry. 
This may have interesting astrophysical and cosmological implications. 

Vice versa, having definite values for the neutrino mixing parameters, it
will be possible, on the basis of requirement of agreement between
the theoretically calculated and the extracted from observations values 
of helium-4, to put cosmological constraints on $L$ even within that range of
exclusively small magnitudes.

Finally we would like to point to the essential differences between our work 
and the previous ones:
we have considered the {\it nonresonant} case of oscillations with very 
{\it small mass 
differences}; the oscillations proceed effectively relatively late i.e. 
after neutrino decoupling and the change in the lepton number is 
dominated {\it by oscillations}, not by collisions; we have accounted 
precisely for the {\it momentum spread} of neutrinos.

Having in mind the latest news from SuperKamioka concerning the 
solar neutrino oscillation solutions~\cite{NU98},
the very small mass differences concerned in our paper may look 
much more attractive, and hence this study will be interesting for 
the solar neutrino audience too. 
 On the other hand we would like to stress that models exploiting 
the suppression of oscillations due to $L$ or the generation of $L$ 
which further suppresses oscillations must be exploited with caution 
and be carefully calculated for the concrete ($L, \delta m^2, 
\vartheta)$ values of interest. As far as it was shown here that the 
effect of asymmetry is not straightforward. Asymmetry - oscillations 
interplay reveals greater complexity than the rough estimations 
suggest and should be studied carefully. 

\section*{Acknowledgements}
We would like to  thank A. Smirnov and G. Senjanovic, B. Bajc 
and E. Akhmedov for useful discussions. We acknowledge the stimulating 
 working atmosphere of the Extended Workshop on Highlights of Astroparticle 
Physics, held in ICTP, November-December 1997, where this work was initiated 
and the hospitality and the financial support of the Abdus Salam
International Center for Theoretical Physics where this work was completed  
during our visiting positions in 1998.    

\pagebreak[1]

\newpage
\onecolumn

\begin{center}{\Large Figure Captions}
\end{center}
\ \\

{\bf Figure 1a}: The evolution of the neutron
number density relative to nucleons $X_n(t)=N_n(t)/(N_p+N_n)$
 for the case of oscillations  with maximal
 mixing and $\delta m^2=10^{-7}$ eV$^2$ for different values of the initial 
asymmetry ( $L=10^{-6}$,  $L=10^{-5}$ and $L=10^{-10}$) is plotted.

\ \\

{\bf Figure 1b}: The evolution of the neutron
number density relative to nucleons $X_n(t)=N_n(t)/(N_p+N_n)$
 for the case of oscillations  with $\sin^2(2\vartheta)=10^{-0.05}$
 and $\delta m^2=10^{-7}$ eV$^2$ for different values of the initial
asymmetry ( $L=10^{-6}$,  $L=10^{-5}$ and $L=10^{-10}$) is plotted.

\ \\

{\bf Figure 2}: The figure illustrates the dependence of the frozen neutron
number density relative to nucleons $X_n=N_n/(N_p+N_n)$ on the
value of the initial neutrino asymmetry for 
$\sin^2(2\vartheta)=10^{-0.05}$, and different mass differences. For comparison 
the standard curve is plotted also with dashed line.
    
\ \\

{\bf Figure 3}: The dependence of the neutron to nucleon freezing ratio 
on the mixing angle for $L=10^{-6}$  for different mass differences 
$\delta m^2 =10^{-7}$ eV$^2$ and $\delta m^2 = 10^{-8}$ eV$^2$    
is shown. For comparison with dashed lines the corresponding curves 
with small asymmetry $L=10^{-10}$ are presented.   

\ \\

{\bf Figure 4}: On the $\delta m^2-\vartheta$ plane the
constant helium contours  calculated in the discussed model of
cosmological nucleosynthesis with neutrino oscillations for $L=10^{-6}$
and $L=10^{-10}$ are shown.

\pagestyle{empty}
\newpage

\epsfbox[50 170 650 770]{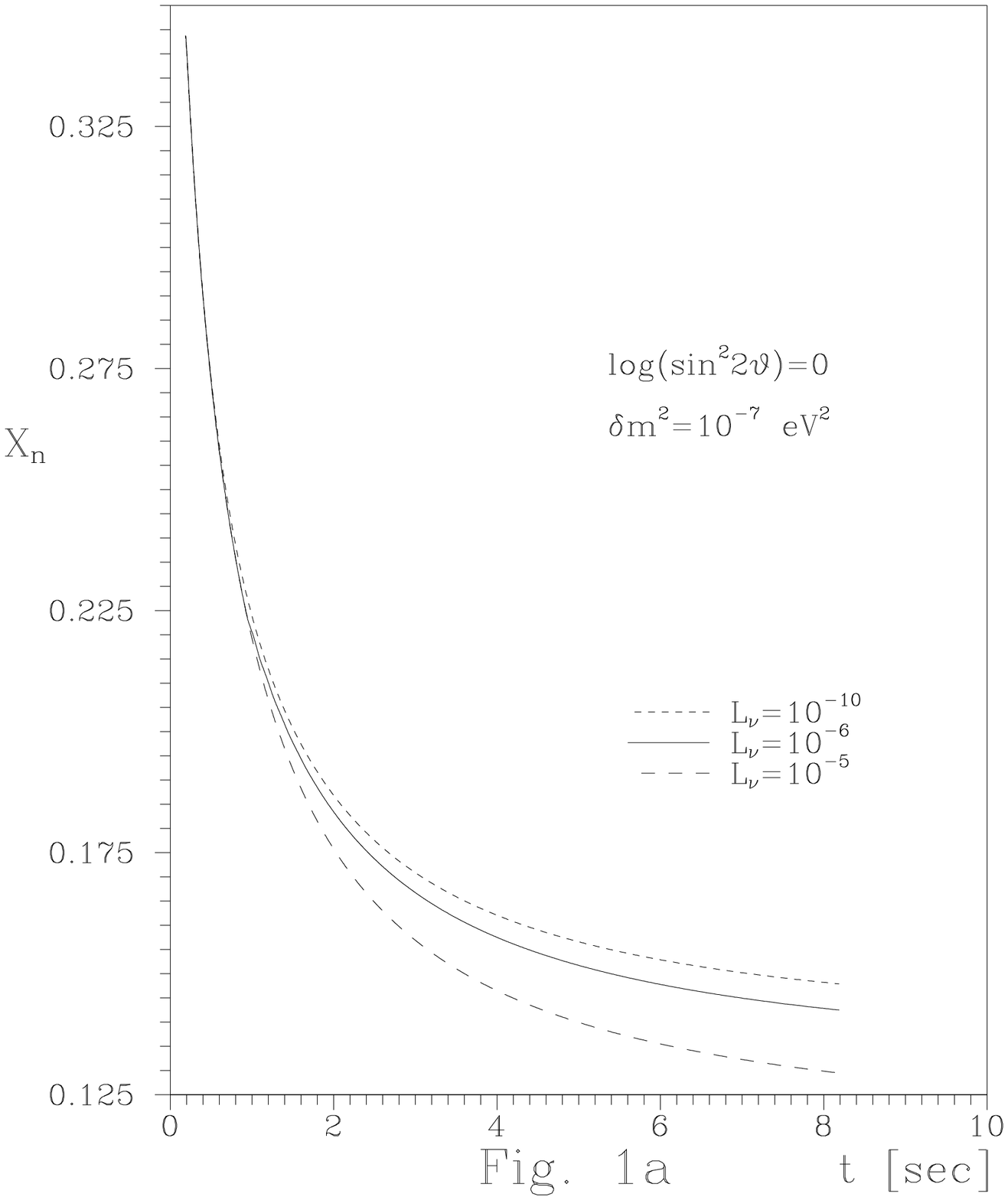}

\vfill

{\bf Figure 1a}: The evolution of the neutron
number density relative to nucleons $X_n(t)=N_n(t)/(N_p+N_n)$
 for the case of oscillations  with maximal
 mixing and $\delta m^2=10^{-7}$ eV$^2$ for different values of the initial 
asymmetry ( $L=10^{-6}$,  $L=10^{-5}$ and $L=10^{-10}$) is plotted.

\newpage

\epsfbox[50 170 650 770]{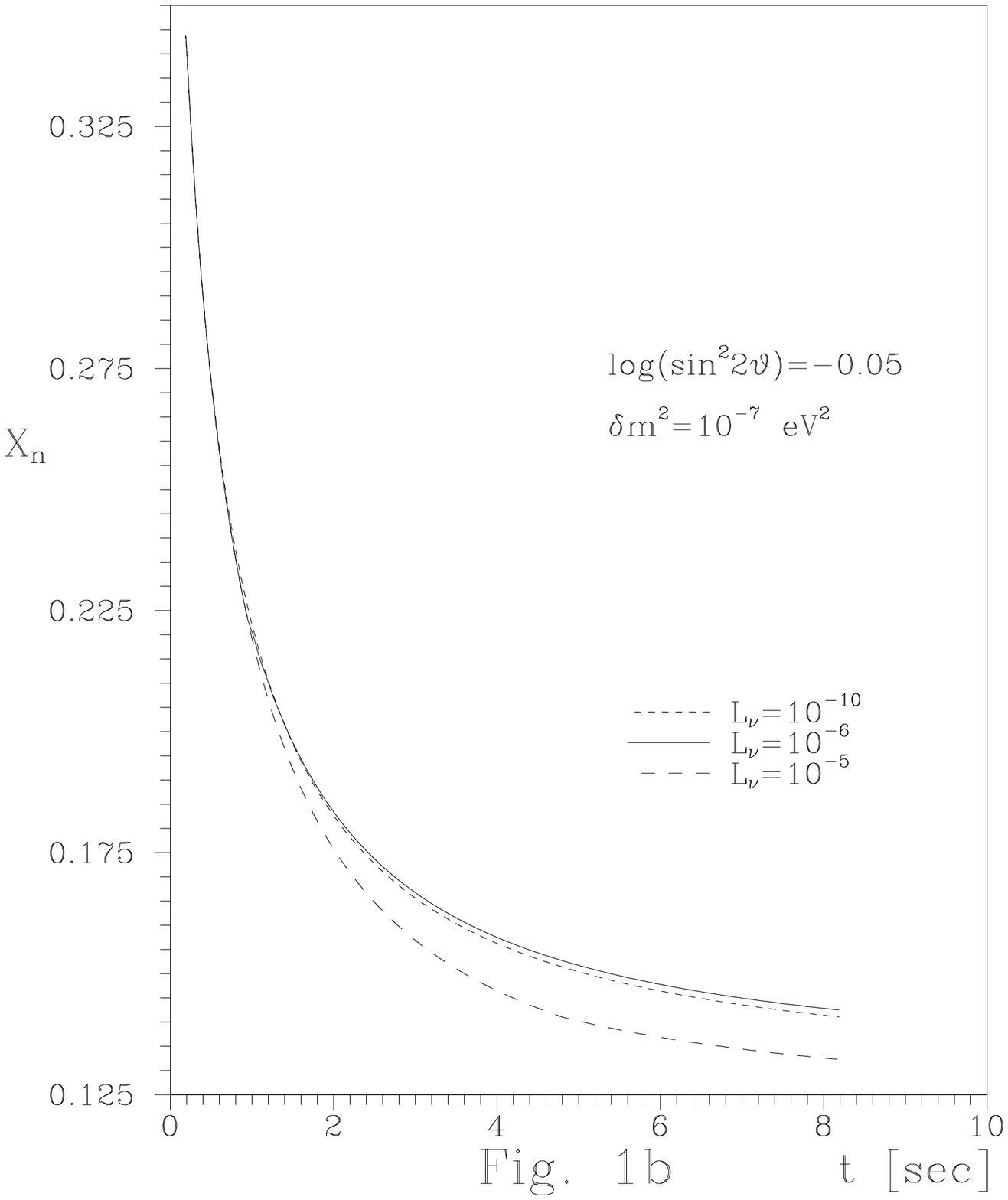}

\vfill

{\bf Figure 1b}: The evolution of the neutron
number density relative to nucleons $X_n(t)=N_n(t)/(N_p+N_n)$
 for the case of oscillations  with $\sin^2(2\vartheta)=10^{-0.05}$
 and $\delta m^2=10^{-7}$ eV$^2$ for different values of the initial
asymmetry ( $L=10^{-6}$,  $L=10^{-5}$ and $L=10^{-10}$) is plotted.

\newpage

\epsfbox[50 170 650 770]{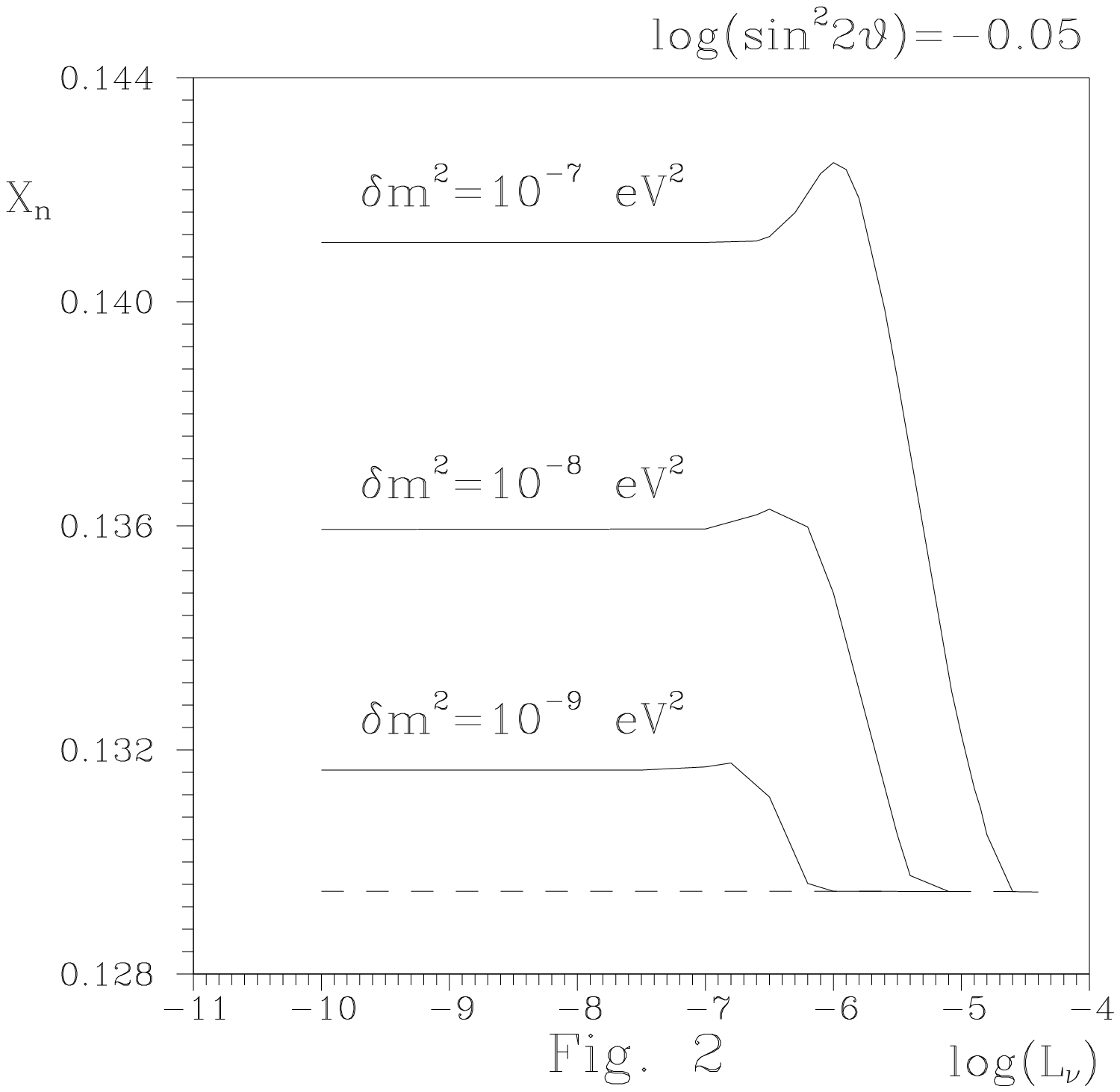}

\vfill

{\bf Figure 2}: The figure illustrates the dependence of the frozen neutron
number density relative to nucleons $X_n=N_n/(N_p+N_n)$ on the
value of the initial neutrino asymmetry for $\sin^2(2\vartheta)=10^{-0.05}$, 
and different mass differences. For comparison 
the standard curve is plotted also with dashed line.

\newpage

\epsfbox[50 170 650 770]{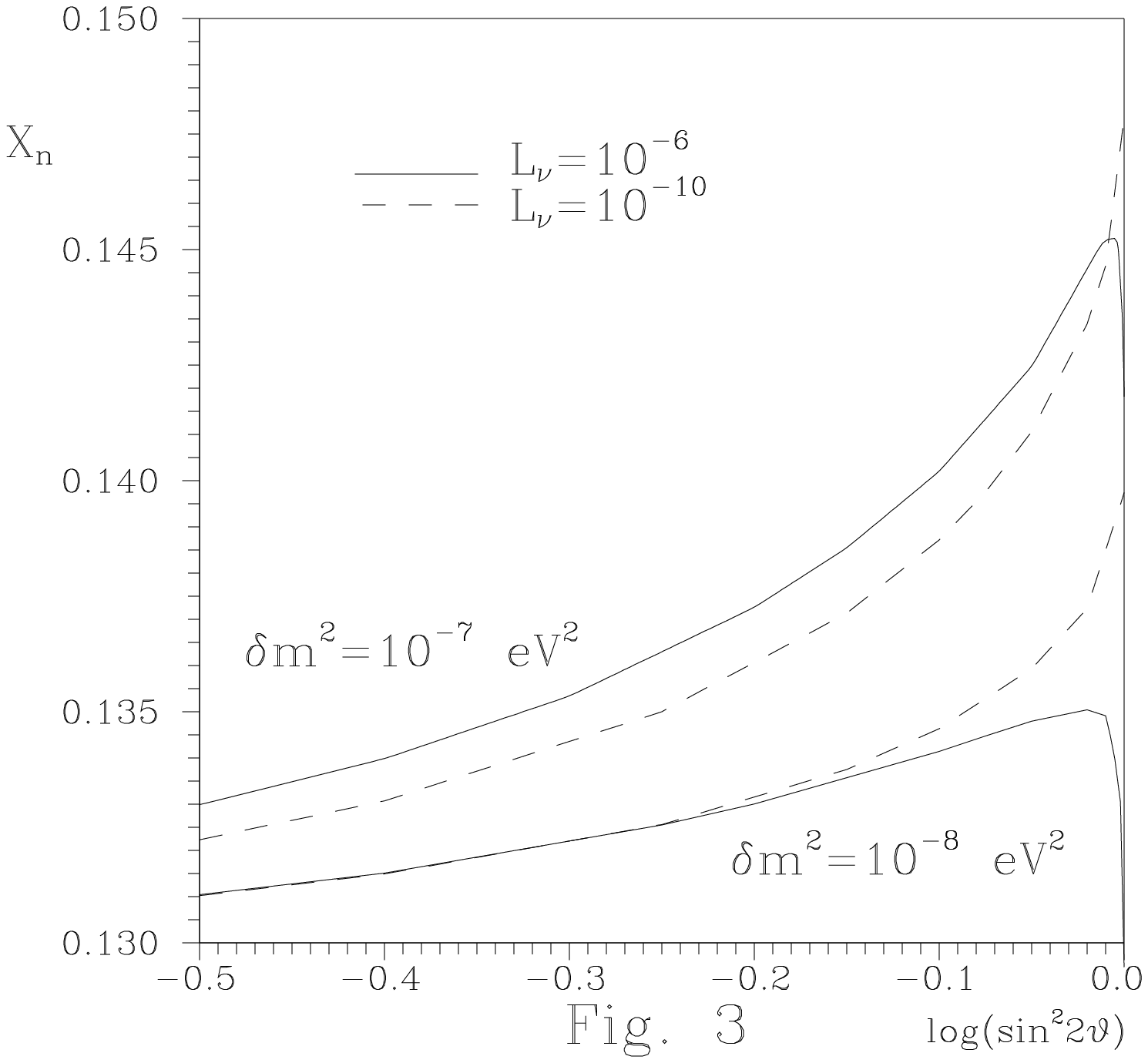}

\vfill

{\bf Figure 3}: The dependence of the neutron to nucleon freezing ratio 
on the mixing angle for $L=10^{-6}$  for different mass differences 
$\delta m^2 =10^{-7}$ eV$^2$ and $\delta m^2 = 10^{-8}$ eV$^2$    
is shown. For comparison with dashed lines the corresponding curves 
with small asymmetry $L=10^{-10}$ are presented.

\newpage

\epsfbox[50 170 650 770]{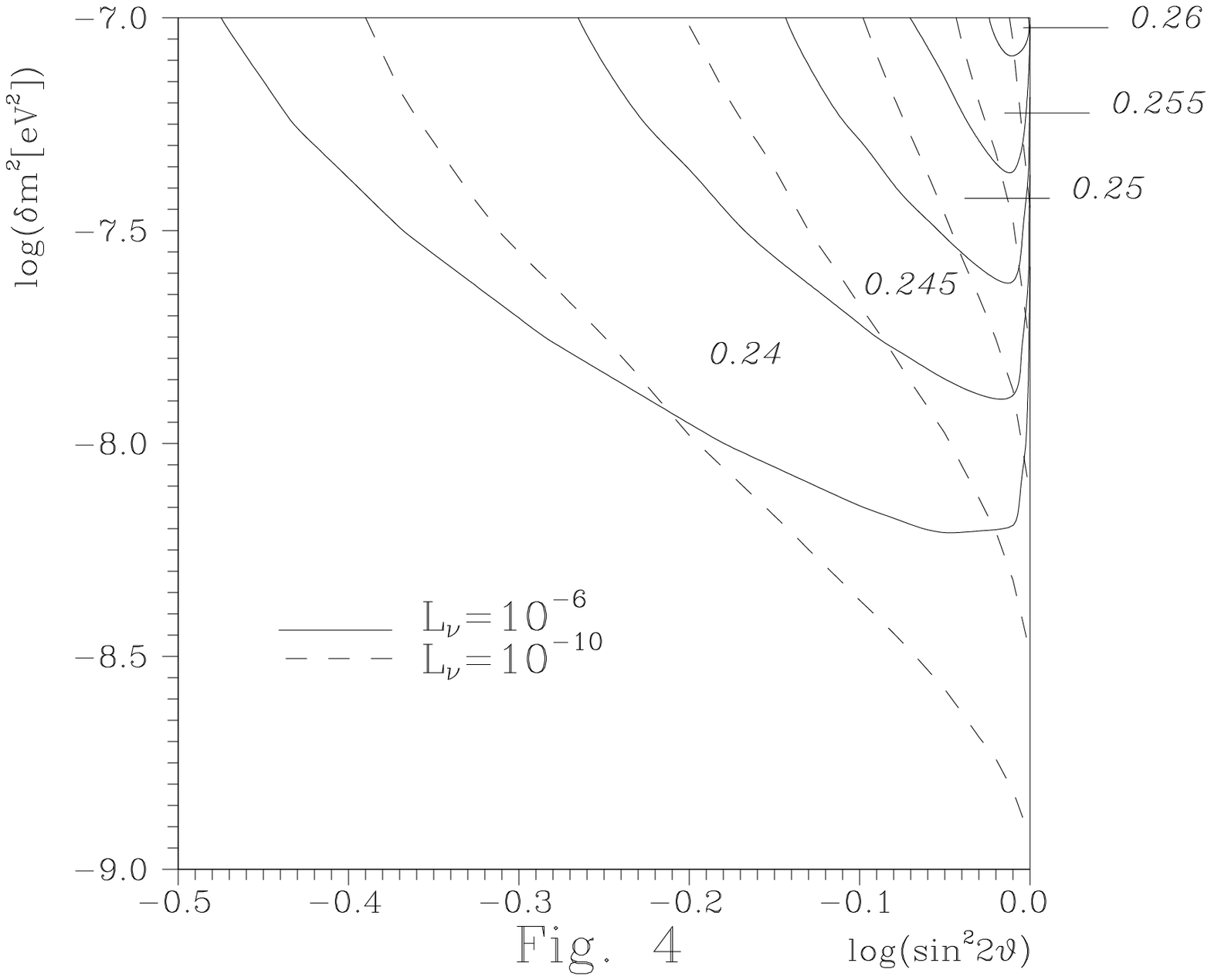}

\vfill

{\bf Figure 4}: On the $\delta m^2-\vartheta$ plane the
constant helium contours  calculated in the discussed model of
cosmological nucleosynthesis with neutrino oscillations for $L=10^{-6}$
and $L=10^{-10}$ are shown.

\end{document}